\documentstyle[aps,prl,preprint]{revtex}
\input epsf

\newcommand{\cb}{\mbox{$\Xi^{0} \rightarrow
 \Sigma^{+}\, e^{-}\, \overline{\nu} $}}

\newcommand{\spls}{\mbox{$\Sigma^{+}$}}
\newcommand{\spf}{\mbox{${\mathbf{P}_{b}}$}}

\newcommand{\fisq}{\mbox{$ f_{1}^{2}$}}
\newcommand{\gisq}{\mbox{$ g_{1}^{2}$}}
\newcommand{\fjsq}{\mbox{$ f_{2}^{2}$}}
\newcommand{\gjsq}{\mbox{$ g_{2}^{2}$}}
\newcommand{\fifj}{\mbox{$ f_{1} f_{2}$}}
\newcommand{\gigj}{\mbox{$ g_{1} g_{2}$}}
\newcommand{\figi}{\mbox{$ f_{1} g_{1}$}}
\newcommand{\figj}{\mbox{$ f_{1} g_{2}$}}
\newcommand{\fjgi}{\mbox{$ f_{2} g_{1}$}}
\newcommand{\fjgj}{\mbox{$ f_{2} g_{2}$}}
\begin{document}
\draft
\title{Effective Hamiltonian Approach to Hyperon Beta Decay with Final-State
Baryon Polarization }

\author{ S. Bright and R. Winston}
\address{The Enrico Fermi Institute and the Department of Physics, \\
The University of Chicago,
Chicago, Illinois 60637}

\author{E. C. Swallow  }
\address{ Department of Physics, Elmhurst College,
 Elmhurst, Illinois 60126 and \\
The Enrico Fermi Institute, The University of Chicago, 
Chicago, Illinois 60637}

\author{A. Alavi-Harati }
\address{ University of Wisconsin, Madison, Wisconsin, 53706 }

\date{\today}
\maketitle
\begin{abstract}

Using an effective Hamiltonian approach, we obtain expressions 
for hyperon $\beta $ decay
final-state baryon polarization.  Terms through 
second order in the energy release
 are retained.  The resulting
approximate expressions are much simpler and more compact
 than the exact expressions, and
they agree closely with them.

\end{abstract}
\pacs{PACS numbers: 13.30.Ce}

In decays such as $\Xi^{-} \rightarrow \Lambda\, e^{-}\, \overline{\nu} $  
or the recently observed\cite{cb} 
$\Xi^{0} \rightarrow \Sigma^{+}\, e^{-}\, \overline{\nu}, $ 
the decay form factors can be probed by observing the
 parity-violating polarization of the 
final-state baryon {\em via} its two body decay mode
 ($\Lambda  \rightarrow p\, \pi^{-}$ or
$\Sigma^{+}  \rightarrow p\, \pi^{0} $ respectively). 
In addition, other kinematic distributions
can be evaluated in the rest frame of the final baryon.

Early analyses of hyperon $\beta$ decay with the 
final-state polarization observed were restricted to 
the zero-recoil approximation \cite{alles}
or were limited in scope\cite{desai}.  More recent
detailed treatments exist \cite{lin,gar}, but the resulting expressions
are quite opaque, and, as a result, the physical content is hidden.  
Also, experiments are not likely to require the exact formulae within 
the foreseeable future.

Using a method introduced by Primakoff for muon capture \cite{pri1,pri2},
we keep only terms through second order in the recoil velocity of the 
initial baryon (in the rest frame of the final baryon).  A similar 
approach has been used to derive
expressions for the case of a polarized initial baryon \cite{ww}.

The most general V-A transition
matrix element for the generic hyperon $\beta $ decay process 
$ B \rightarrow b\, e^{-}\, \overline{\nu} $ can 
be written \cite{bj} in the form

\begin{equation}
{\cal M } =  G_{S}\frac{\sqrt{2}}{2} \overline{u}_{b}  
( O_{\alpha}^{V} + O_{\alpha}^{A} )
u_{B} \overline{u_{e}} \gamma^{\alpha} (1+ \gamma_{5} )
v_{\nu}  + H.c. ,
\end{equation}

where 


\begin{eqnarray}
 O_{\alpha}^{V} & = & f_{1} \gamma_{\alpha} 
+ \frac{f_{2}}{M_{B}}\sigma_{\alpha \beta} 
q^{\beta} +  \frac{f_{3}}{M_{B}}q_{\alpha}, \nonumber \\
 O_{\alpha}^{A} &  = & ( g_{1} \gamma_{\alpha} + \frac{g_{2}}{M_{B}}
\sigma_{\alpha \beta} 
q^{\beta} +  \frac{g_{3}}{M_{B}}q_{\alpha} ) \gamma_{5}, \nonumber \\
q^{\alpha} & = & ( p_{e} + p_{\nu} )^{\alpha} 
= ( p_{B} - p_{b} )^{\alpha},\  \  \mathrm{and}
\end{eqnarray}

\[ G_{S} = \left\{ \begin{array}{ll}
        G_{F} V_{us}  &    for \mid \Delta S \mid  = 1 \\
        G_{F} V_{ud}  &    for\  \  \Delta S  = 0. \\
     \end{array}
\right. \]

Here ${G_{F}} $ is the Fermi coupling constant, $ V_{us} $ and $ V_{ud} $
are the appropriate Cabibbo-Kobayashi-Maskawa matrix elements,
and $\Delta S$ denotes the
strangeness change in the decay.  

We relate the transition matrix element to an effective Hamiltonian by
\begin{equation}
{\cal M } =  \langle be \mid {\cal H_{\mbox{eff}}} \mid B \nu \rangle 
\sqrt{2e\,2\nu\,2M_{b}\,(E_{B}+M_{B})}
\end{equation}
with
\begin{eqnarray}
\frac{\sqrt{2}}{2}{\cal H_{\mbox{eff}}} & = & 
G_{S} \: \frac{1}{2}(1 - {\mathbf{\sigma}_{\ell}} \cdot \hat{e} ) \,
[G_{V} + G_{A} {\mathbf{\sigma}_{\ell}} 
\cdot  {\mathbf{\sigma}_{b}} \nonumber \\
&  & + G_{P}^{e} {\mathbf{\sigma}_{b}} \cdot \hat{e} 
+ G_{P}^{\nu} {\mathbf{\sigma}_{b}} \cdot \hat{\nu} ] \,
\frac{1}{2}(1 - {\mathbf{\sigma}_{\ell}} \cdot \hat{\nu} ).
\end{eqnarray}
Here $\hat{e}$ and $\hat{\nu}$ are unit vectors along the
electron and antineutrino directions, while $e$, $\nu$, and $E_{B} $ are
the energies of the electron, antineutrino,
and initial baryon (all quantities are in the rest frame of $b$).
The spin operators $ {\mathbf{\sigma}_{\ell}} $ and $ {\mathbf{\sigma}_{b}}$ 
act respectively on the
lepton and baryon states (represented by two-component spinors).

The effective coupling coefficients 
$G_{V}$, $ G_{A}$, $G_{P}^{e}$, and $G_{P}^{\nu}$ are
functions of the form factors in eqn. (2): 
\begin{eqnarray}
G_{V} & =  & f_{1} + \delta f_{2} - \frac{\nu+e}{2M_{B}}
(  f_{1} + \Delta f_{2} ),  \nonumber \\
G_{A} & = & -g_{1} + \delta g_{2} + \frac{\nu-e}{2M_{B}}
(  f_{1} + \Delta f_{2} ),  \nonumber \\
G_{P}^{e} & = & \frac{e}{2M_{B}}( - ( f_{1} + \Delta f_{2} )
 - g_{1} + \Delta g_{2} ),  \nonumber \\
G_{P}^{\nu} & = & \frac{\nu}{2M_{B}}( f_{1} + \Delta f_{2} 
 - g_{1} + \Delta g_{2} ),  
\end{eqnarray}
where $ \delta = (  M_{B} -   M_{b} ) / M_{B} $ 
and $ \Delta = (  M_{B} +   M_{b} ) / M_{B} = 2 - \delta $.
Since the form factors $  f_{3} $ and $ g_{3} $ 
always appear with a multiplier of the electron mass
 divided by $M_B$, they are neglected
throughout.  Note also that $  f_{2} $ and $ g_{2} $
 always appear multiplied by a quantity of
order  $ \delta $, so their $ q^{2}$ dependence is not
 relevant to our order $ \delta^{2} $
approximation.  However, the $ q^{2}$ dependence of
 $ f_{1} $ and $ g_{1} $ does need to be
included\cite{gar} in calculations to maintain a completely
 consistent order of approximation.

Electron and antineutrino spins are not usually observed, and 
this analysis focuses on measurement of the final baryon 
polarization.  We therefore 
sum over the electron and antineutrino spins and average over
initial baryon spin:  

\begin{equation}
\sum_{\nu \: {\rm spins}, B \: {\rm spins} }^{}
\mid \langle be\!\mid { \cal H_{\mbox{eff}}  } \mid B \nu \rangle \mid^{2} =
\langle be\!\mid { \cal H_{\mbox{eff}} H_{\mbox{eff}}^{\dagger}  } 
\mid be \rangle 
\end{equation}
and
\begin{equation}
\sum_{e \: {\rm spins}  }^{}
\langle be\!\mid { \cal H_{\mbox{eff}} H_{\mbox{eff}}^{\dagger}  }
 \mid be \rangle =
Tr((1+{\bf \sigma_{b}}\cdot \spf){ \cal H_{\mbox{eff}} 
H_{\mbox{eff}}^{\dagger}} ).
\end{equation}

By projecting out the spin of the final baryon and taking the trace, 
we obtain

\begin{eqnarray}
\mid {\cal M } \mid^{2} &  = & \xi[1+a \hat{e} \cdot \hat{\nu} +
A \spf \cdot \hat{e} +  B \spf \cdot \hat{\nu}  \nonumber \\
& & 
+ A' ( \spf \cdot \hat{e} ) ( \hat{e} \cdot \hat{\nu} )
+ B' ( \spf \cdot \hat{\nu} ) ( \hat{e} \cdot \hat{\nu} )  
\nonumber \\
& &
+ D \spf \cdot ( \hat{e} \times  \hat{\nu} ) ] \nonumber \\
& & 
\cdot (2e)(2\nu)(2M_{b})(E_{B} + M_{B})G_{S}^{2},
\nonumber \\
\xi & = & \mid\!G_{V}\!\mid^{2} + 3  \mid\!G_{A}\!\mid^{2} 
 - 2 Re( G_{A}^{*}( G_{P}^{e} +  G_{P}^{\nu} )) \nonumber \\
& & 
+ \mid\!G_{P}^{e}\!\mid^{2}
+ \mid\!G_{P}^{\nu}\!\mid^{2}, \nonumber \\
\xi a & = & \mid\!G_{V}\!\mid^{2} - \mid\!G_{A}\!\mid^{2} 
- 2 Re( G_{A}^{*}( G_{P}^{e} +  G_{P}^{\nu} ))\nonumber \\
& &
+ \mid\!G_{P}^{e}\!\mid^{2}
+ \mid\!G_{P}^{\nu}\!\mid^{2} 
+ 2 Re( G_{P}^{e*} G_{P}^{\nu} )( 1
 + \hat{e} \cdot \hat{\nu} ), \nonumber \\
\xi A & = & - 2 Re( G_{V}^{*} G_{A} ) + 2  \mid\!G_{A}\!\mid^{2}  \nonumber \\
& & + 2 Re( G_{V}^{*} G_{P}^{e} -  G_{A}^{*} G_{P}^{\nu} ), \nonumber \\
\xi B & = & - 2 Re( G_{V}^{*} G_{A} ) - 2  \mid\!G_{A}\!\mid^{2}  \nonumber  \\
& & + 2 Re( G_{V}^{*} G_{P}^{\nu} + G_{A}^{*} G_{P}^{e} ), \nonumber \\
\xi A' & = & 2 Re( G_{P}^{e*} (  G_{V} - G_{A} ) ), \nonumber \\
\xi B' & = & 2 Re( G_{P}^{\nu*} (  G_{V} + G_{A} ) ), \nonumber \\
\xi D  & = & 2 Im(  G_{V}^{*} G_{A} ) + 2 Im(  G_{P}^{e*} G_{P}^{\nu} ) 
( 1 +  \hat{e} \cdot \hat{\nu} ) \nonumber \\
& &  + 
 2 Im( G_{A}^{*} (  G_{P}^{e} -  G_{P}^{\nu} ) ). 
\end{eqnarray}

The polarization of the final baryon may be expressed explicitly as

\begin{equation}
\spf = \frac{ (A + A' \hat{e} \cdot \hat{\nu} ) \hat{e}
+ (B + B' \hat{e} \cdot \hat{\nu} ) \hat{\nu}
+ D  \hat{e} \times  \hat{\nu} }{ 1 + a \hat{e} \cdot \hat{\nu} }.
\end{equation}
The components of this polarization can
readily be measured when the outgoing 
baryon $b$ is a hyperon which undergoes a subsequent 
weak decay $ b  \rightarrow b' \pi $ with a non-zero decay 
asymmetry parameter $ \alpha_{b'} $.  The distribution of the $ b' $ 
direction relative to any axis defined by a unit 
vector $ \hat{i} $ is given by 

\begin{equation}
\frac{1}{\Gamma} \;  \frac{d \Gamma }{d \Omega_{b'}} =
\frac{1}{4 \pi }( 1 + {\mathsf{S}_{i}} \alpha_{b'} \hat{i} \cdot \hat{b'} ),
\end{equation}
where $ {\mathsf{S}_{i}} =  \langle {\mathbf{P}_{b}} \cdot \hat{i} \rangle $ 
is the average 
polarization of $b$ in the $ \hat{i}$ direction.
Conceptually, it is advantageous to employ the orthonormal basis

\begin{eqnarray}
\hat{\alpha} & = & 
\frac{  \hat{e} + \hat{\nu} }{ \sqrt{ 2(1 
+ \hat{e} \cdot \hat{\nu})}}, \nonumber \\
\hat{\beta}  & = & 
\frac{  \hat{e} - \hat{\nu} }{ \sqrt{ 2(1 
- \hat{e} \cdot \hat{\nu})}}, \nonumber \\
\hat{\gamma} & = &  \hat{\alpha} \times \hat{\beta}.
\end{eqnarray}
Experimentally, it may be more advantageous to
 determine the polarization components along
one or more of the outgoing particle
 directions ($ \hat{e},  \hat{\nu}, \hat{b} $).

To gauge the importance of the recoil contributions, in Fig 1 
we compare values of several integrated observables calculated from our 
expressions with the corresponding zero-recoil values for the decay 
$\Xi^{0} \rightarrow \Sigma^{+} \, e^{-} \, \overline{\nu} $.  
For these calculations, we assumed $ V_{us} = 0.2205 $, $ f_{1}(0) = 1.0 $, 
$ f_{2} = 2.6 $, and $ g_{2} = 0.0 $.
Comparing values of integrated observables obtained from our expressions 
with exact values from tables in Ref.\cite{gar}, 
we find that the decay rates agree to better than 1 \%, and that 
polarizations and 
asymmetries agree to better than 0.004, the differences being almost entirely
due to terms of order $\delta^3$.  We have not 
included electromagnetic corrections,
which are discussed in Ref.\cite{gar}.

Finally, we present the analytic expressions for the integrated
final state polarization
to order $ \delta $ in the final state rest frame, 
assuming real form factors.  The order $ \delta^2 $ expressions
can be obtained by adding the $ {\cal O}( \delta^2 ) $ terms given in 
\cite{snd}.

\begin{eqnarray}
R & = &  R_{0} [ ( 1 - \frac{3}{2} \delta ) \fisq
      +  ( 3 - \frac{9}{2} \delta ) \gisq 
      -  ( 4 \delta ) \gigj ] + R(\delta^2) , \nonumber \\
R \mathsf{S}_{e}  & = &  R_{0} [ ( 2 - \frac{10}{3} \delta ) \gisq
      +  ( 2 - \frac{7}{3} \delta ) \figi
      -  (\frac{1}{3} \delta ) \fisq \nonumber  \\
& &
      -  (\frac{2}{3} \delta ) \fifj 
      +  (\frac{2}{3} \delta ) \fjgi
      -  (\frac{2}{3} \delta ) \figj
      -  (\frac{10}{3} \delta  ) \gigj ]
      + R \mathsf{S}_{e} (\delta^2) , \nonumber \\
R  \mathsf{S}_{\nu} & = &  R_{0} [( -2 + \frac{10}{3} \delta ) \gisq
      +  ( 2 - \frac{7}{3} \delta ) \figi
      +  ( \frac{1}{3} \delta ) \fisq \nonumber  \\
& &
      +  ( \frac{2}{3} \delta ) \fifj 
      +  ( \frac{2}{3} \delta ) \fjgi
      -  ( \frac{2}{3} \delta ) \figj
      +  ( \frac{10}{3} \delta ) \gigj ]
      + R \mathsf{S}_{\nu} (\delta^2) , \nonumber \\
R \mathsf{S}_{\alpha}  & = & R_{0}
[( \frac{8}{3} - \frac{52}{15} \delta ) \figi
      +  (\frac{16}{15} \delta ) \fjgi
      -  (\frac{16}{15} \delta ) \figj ]
      + R \mathsf{S}_{\alpha} (\delta^2), \nonumber \\
R \mathsf{S}_{\beta}  & = &  R_{0} [( \frac{8}{3} - 4 \delta )\gisq
      -  (\frac{8}{15} \delta ) \fisq
      -  (\frac{16}{15} \delta ) \fifj  \nonumber \\
& &
      -  (\frac{64}{15} \delta ) \gigj ]
      + R \mathsf{S}_{\beta} (\delta^2), 
\end{eqnarray}

where 

\begin{eqnarray*}
R_{0} = \frac{G_{S}^{2} (\delta M_{B})^{5}}{60 \pi^{3}}. \nonumber
\end{eqnarray*}

As can be seen in Ref. \cite{alles}, the zero-recoil $( \delta =
0)$ expression for $ \mathsf{S}_{e} (\mathsf{S}_{\nu} $) 
is the same as that for
the neutrino (electron) asymmetry for a polarized
initial baryon \cite{gar}.  Also, $ R \mathsf{S}_{\alpha}$ depends only on
$ V \times A $ cross terms, and $ R \mathsf{S}_{\beta}$
 depends only on $ V \times V $ and
$ A \times A $ terms, as required by a theorem due to Weinberg
\cite{sw}.

We thank J.L.~Rosner for helpful comments and discussions.  
The continuing stimulation
provided by our colleagues in the KTeV collaboration
at Fermilab, especially 
T.~Alexopoulos, A.R.~Erwin, D.A.~Jensen
, E.~Monnier, E.J.~Ramberg, N.~Solomey and M.~Timko, is also gratefully
acknowledged.  This work was supported in part
 by the U. S. Department of Energy under grants
DE-FG02-90ER40560 (Task B) and DE-FG02-95ER40896 (Task A).
One of us (SB) acknowledges receipt of a U. S. Department 
of Education GANN Fellowship.



\onecolumn
\begin{figure}
\epsfxsize=20.cm       
\epsfbox{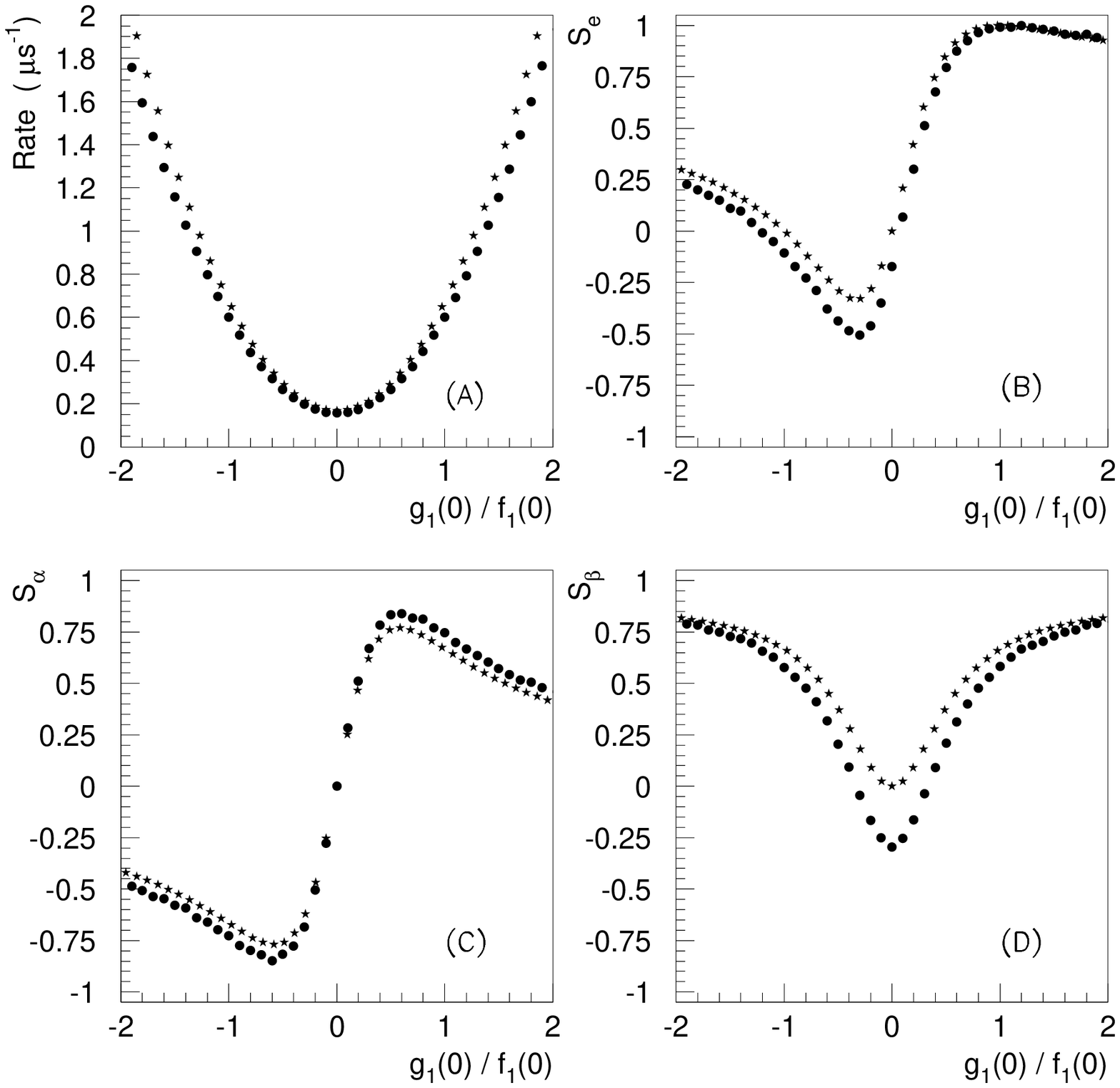}
\caption{Integrated observable quantities for the decay \cb \, as a function 
of $ g_{1} / f_{1} $ : \\
A) The total decay rate $ ( \mu s^{-1} ) $; \\
B) The polarization of the \spls in the $ e^{-} $  
direction $ ({\mathsf{S}_{e}} =  
\langle {\mathbf{P}_{b}} \cdot \hat{e} \rangle ) $; \\
C) The polarization of the \spls in the $ \alpha $ 
direction $ ({\mathsf{S}_{\alpha}} =  
\langle {\mathbf{P}_{b}} \cdot \hat{\alpha} \rangle ) $; \\
D) The polarization of the \spls in the $ \beta $  
direction $ ({\mathsf{S}_{\beta}} =  
\langle {\mathbf{P}_{b}} \cdot \hat{\beta} \rangle ) $. \\
The stars ( $\star$ ) are zero recoil values, and circles ( $\bullet$ ) are  
values obtained by numerical integration of our formulae ( correct to 
order $\delta^2$ ).}
\label{fig:int}
\end{figure}


\begin{references}
\bibitem{snd} The correct order $\delta^2$ expressions are obtained by adding
\begin{eqnarray*}
R(\delta^2) & = &   R_{0} \delta^2 (\frac{6}{7} \fisq
+ \frac{12}{7} \gisq + 6 \gigj  \nonumber \\
& &
+ \frac{6}{7} \fifj + \frac{4}{7} \fjsq + \frac{12}{7} \gjsq ),\nonumber \\
%
R \mathsf{S}_{e}(\delta^2) & = &  R_{0} \delta^2 ( \frac{55}{42} \gisq
+ \frac{17}{21} \figi + \frac{19}{42} \fisq
+ \frac{4}{3} \fifj - \frac{10}{21} \fjgi \nonumber \\
& & 
+ \frac{10}{21} \figj + \frac{116}{21} \gigj 
+ \frac{4}{21} \fjsq + \frac{4}{3} \gjsq -  \frac{16}{21} \fjgj ), \nonumber \\
%
R \mathsf{S}_{\nu}(\delta^2) & = &  R_{0} \delta^2 ( -\frac{55}{42} \gisq
+  \frac{17}{21} \figi - \frac{19}{42} \fisq
- \frac{4}{3} \fifj - \frac{10}{21} \fjgi \nonumber \\
& &
+ \frac{10}{21} \figj - \frac{116}{21} \gigj 
-  \frac{4}{21} \fjsq - \frac{4}{3} \gjsq -  \frac{16}{21} \fjgj ), \nonumber \\
%
R \mathsf{S}_{\alpha}(\delta^2) & = &  R_{0} \delta^2 ( \frac{316}{245} \figi
- \frac{752}{735} \fjgi + \frac{752}{735} \figj 
- \frac{128}{105} \fjgj  ), \nonumber \\
%
R \mathsf{S}_{\beta}(\delta^2) & = &  R_{0} \delta^2 ( 
\frac{422}{735} \fisq + \frac{88}{49} \fifj + \frac{8}{35} \fjsq  \nonumber \\
& &
+ \frac{362}{245} \gisq + \frac{1576}{245} \gigj + \frac{8}{5} \gjsq )
\end{eqnarray*}
to $ R, R\mathsf{S}_{e}, R\mathsf{S}_{\nu}, R\mathsf{S}_{\alpha} $ and
$ R\mathsf{S}_{\beta}$, respectively in equation 12.
\end{references}
\end{document}